\documentclass{jetpl}
\usepackage{epsfig,amssymb,amsfonts,bm,slashbox}
\twocolumn

\DeclareMathSymbol{\varSigma}{\mathord}{letters}{"06}
\newcommand{\be}{\begin{equation}}
\newcommand{\ee}{\end{equation}}
\newcommand{\ds}{\displaystyle}

\newcommand{\vep}{{\bm p}}

\newcommand{\veq}{{\bm q}}
\newcommand{\ves}{{\bm s}}
\newcommand{\veg}{{\bm g}}

\newcommand{\bea}{\begin{eqnarray}}
\newcommand{\eea}{\end{eqnarray}}
\newcommand{\beas}{\begin{eqnarray*}}
\newcommand{\eeas}{\end{eqnarray*}}

\renewcommand{\thetable}{\arabic{table}}

\title{One-pion exchange in the $X(3872)$ revisited}
\rtitle{One-pion exchange in the $X(3872)$ revisited}

\author{Yu.\,S.\,Kalashnikova}
\address{Institute of Theoretical and Experimental Physics,
117218, B.Cheremushkinskaya 25, Moscow, Russia}

\author{A.\,V.\,Nefediev}
\address{Institute of Theoretical and Experimental Physics,
117218, B.Cheremushkinskaya 25, Moscow, Russia\\
National Research Nuclear University MEPhI, 115409, Moscow, Russia\\
Moscow Institute of Physics and Technology, 141700, Dolgoprudny, Moscow Region, Russia\\
All-Russia Research Institute of Automatics (VNIIA), 127055, Moscow, Russia}

\rauthor{Yu.\,S.\,Kalashnikova, A.\,V.\,Nefediev}

\abstract{We re-examine one-pion exchange as a possible binding mechanism in the $X(3872)$ charmonium-like state and argue it to be not sufficiently binding for this purpose. We conclude therefore that other
short-range dynamics are responsible for the $X$ formation.}


\begin{document}
\maketitle

\setcounter{figure}{0}
\setcounter{table}{0}

\section{Introduction}

A charmonium-like state $X(3872)$ was observed by Belle Collaboration in 2003 \cite{Xobservation} in $B$-meson decays
in the mode $B^+\to K^+X\to K^+\pi^+\pi^- J/\psi$, and since then this state attracts attention of both experimentalists and
theorists due to very specific properties it possesses. This state is also seen in the mode $\pi^+\pi^-\pi^0 J/\psi$
\cite{Babar3pi}, with approximately the same branching fraction as that in the mode $\pi^+\pi^- J/\psi$ (for a recent
update see \cite{Belle2pi}), that clearly indicates a large isospin violation in the $X$. There is no general
agreement on the quantum numbers of this state yet: while the analysis of the $\pi^+\pi^- J/\psi$ decay mode yields
either $J^{PC}=1^{++}$ or $J^{PC}=2^{-+}$ quantum numbers \cite{Belle2pi,rho}, the recent analysis of the $\pi^+\pi^-\pi^0 J/\psi$
mode seems to favour the $2^{-+}$ assignment \cite{Babar3pi}, though the $1^{++}$ option is not excluded. Quite opposite
claims can be found in the literature (see, for example, \cite{Xqn1} versus
\cite{Xqn2}), however there is a general consensus that, if $2^{-+}$ quantum numbers are confirmed, very exotic
theoretical models for the $X$ will have to be invoked to explain all established properties of the latter
\cite{chinese,burns,1D2}. In addition, $X(3872)$ resides exactly at the $D^0\bar{D}^{*0}$\footnote{A proper $C$-parity eigenstate is always meant by this (and similar) shorthand notation.} threshold that, for the
quantum numbers $1^{++}$, can be explained naturally by a strong attraction of a conventional $2^3P_1$ $c\bar{c}$ charmonium
to the $S$-wave $D\bar{D}^*$ threshold \cite{coupled1,coupled2}. Molecule model
\cite{molecule1,molecule2,molecule3,molecule4} is also consistent only with the option $1^{++}$. In this Letter we
stick to the quantum numbers $1^{++}$, as to the most promising candidate, and revisit the long standing problem of the
one-pion exchange (OPE) as a possible binding mechanism responsible for the formation of the $X$.

Pion exchange between charmed mesons was suggested long ago \cite{molecule1,molecule2} as a mechanism able to bind
the isosinglet $D \bar D^{*}$ mesonic system and to form a deuteron-like state near threshold. This model was
revisited shortly after the $X(3872)$ discovery \cite{molecule3,molecule4}, while further implications of the nearby
pion threshold are discussed in~\cite{braatenpions,pions}. According to a combined analysis of the most recent
data for the $X\to\pi^+\pi^- J/\psi$ and $X\to D^0\bar{D}^0\pi^0$ modes
\cite{Braaten:2007dw,Kalashnikova:2009gt,Kalashnikova:2010zz}, $X(3872)$ is a bound state localised within
approximately 1~MeV below the neutral $D\bar{D}^*$ threshold. Therefore a relevant question arises concerning the
mechanisms responsible for the formation of this bound state, and the OPE is a possible candidate for this role. This
problem is addressed in~\cite{ch,ThCl}. In particular, only the neutral
$D^0\bar{D}^{*0}$ configuration is studied in~\cite{ch} and the conclusion made is in negative, namely that the
OPE appears to be unable to bind such a system. However, in~\cite{ThCl}, the charged $D \bar D^*$ channel is
also taken into account and the opposite conclusion is made that the resulting OPE interaction is able
to produce the $X$ as a shallow bound state. Notice that both above-mentioned calculations treat the $D\bar{D}^*$ system
in a
deuteron-like fashion: OPE enters in the form of a static potential. In the meantime, the $D^{*0}$ mass is very close
to the $D^0 \pi^0$ threshold and thus a relevant worry \cite{suzuki} is that in the $D\bar{D}^*$ system, bound by the
OPE, the pion may go on shell. The latter calls for the proper inclusion of the three-body $D\bar{D}\pi$ unitarity
cuts. As shown in~\cite{deeply}, the cut effects are of paramount importance in the
$D_{\alpha}\bar{D}_{\beta}$ system, if the $D_{\beta}$ width is dominated by the $S$-wave $D_{\beta}\to D_{\alpha}\pi$
decay. In particular, (deeply) bound $D_1 \bar{D^*}$
states found in~\cite{closedeeply1,closedeeply2} in the static approximation disappear completely from the
spectrum if the full three-body treatment is invoked. In the case of the $X(3872)$ one deals with the $P$-wave
$D^*\to D\pi$ vertex, so no disastrous consequences are expected from the inclusion of the cut effects
(see~\cite{Baru:2011rs} for a detailed discussion of the role played by the three-body dynamics in the $X(3872)$),
however the problem has to be treated with lots of caution because of the divergent $D$-meson loop integrals. In
\cite{ch,ThCl} divergent integrals are made finite with the help of form factors of a suitable form, with a
cut-off parameter $\Lambda$, and the conclusion whether the OPE interaction is binding enough is made as based on the
value of $\Lambda$ necessary to produce a bound state with the given binding energy. In particular, it is shown in
\cite{ch} that a bound state in the $D^0\bar{D}^{*0}$ system, with the binding energy around 1~MeV, exists only for
the values of $\Lambda$ of order of 6~GeV, that is for values much larger than those admitted by interpretation of the
form factors in terms of the quark model. On the contrary, it is argued in \cite{ThCl} that, already for as small
cut-off's as just 1~GeV, a bound state with the binding energy 1~MeV appears, if the charged channel is taken into
account (and provides an extra attraction in the system). This result is interpreted then in \cite{ThCl} as a proof that
the OPE provides enough attraction to produce a bound state.

In our research we also adopt the above mentioned regularisation scheme, that is we introduce a form factor with the
cut-off $\Lambda$, interpret values of $\Lambda$ below 1~GeV as phenomenologically adequate and disregard larger values
of $\Lambda$ as unphysical. Indeed larger values of the cut-off cannot be justified in the framework of quark models.
Then, in our model we (i) include both neutral and charged $D\bar{D}^*$ channels, (ii) go beyond the static
approximation used in \cite{ch,ThCl} and include the imaginary part of the OPE into consideration, (iii) treat pions as
relativistic particles, but stick to nonrelativistic dynamics for the $D$ and $D^*$ mesons, as prompted by the values of
the loop momentum of order of the cut-off $\Lambda\lesssim 1$~GeV.

The result of this improved calculation which we report
here is that no bound state exists in the system for $\Lambda\sim$ 1~GeV, and
it appears at threshold for as large values of the cut-off as $\Lambda>2$~GeV. Thus
the conclusion we are led to is that OPE apparently fails to be sufficiently binding to produce the $X$ as a
bound state. Therefore different, short-range, mechanisms are responsible for the $X$ formation.

\section{$D\bar{D}^*$ OPE potential}

We start from the discussion of the $P$-wave $D^*\to D \pi$ vertex which can be parametrised
in the form
\be
v_\mu=g_f\bar {u}^*_{\alpha}(\tau^a)^{\alpha}_{\beta}u^{\beta}\pi^a p_{\pi\mu},
\label{fvertex}
\ee
where $p_{\pi\mu}$ is the pion 4-momentum, $u^*$, $u$, and $\pi$ are the isospin wave functions of the $D^*$, $D$, and
the pion, respectively. Then various $D^*\to D \pi$ decay widths can be evaluated as
\bea
&&\varGamma(D^{*0}\to D^0\pi^0)=\frac{g_f^2 q_{00}^3}{24\pi m_{*0}^2},
\label{width0}\\
&&\varGamma(D^{*+}\to D^+\pi^0)=\frac{g_f^2 q_{c0}^3}{24\pi m_{*c}^2},
\label{widthc}\\
&&\varGamma(D^{*+}\to D^0\pi^+)=\frac{g_f^2 q_{0c}^3}{12\pi m_{*c}^2},
\eea
where $q_{00}$, $q_{c0}$, and $q_{0c}$ are the $D^0$-$\pi^0$, $D^+$-$\pi^0$, and $D^0$-$\pi^+$ relative momentum,
respectively. Notice that, up to a kinematical factor, $\varGamma(D^{*+}\to D^0\pi^+)$ is two times larger than 
$\varGamma(D^{*0}\to D^0\pi^0)$ and $\varGamma(D^{*+}\to D^+\pi^0)$.

Here and in what follows $m_0$, $m_c$, $m_{*0}$, $m_{*c}$, $m_{\pi^0}$, and $m_{\pi^c}$ are the masses of the neutral
and charged $D$ mesons, $D^*$ mesons, and pions, respectively. In our calculations we use the following values
\cite{PDG}:
\begin{eqnarray*}
m_{\pi^0}=134.98~\mbox{MeV},\quad m_{\pi^c}=139.57~\mbox{MeV},\\
m_0=1864.84~\mbox{MeV},\quad m_c=1869.62~\mbox{MeV},\\
m_{*0}=2006.97~\mbox{MeV},\quad m_{*c}=2010.27~\mbox{MeV}.
\end{eqnarray*}

To have a better contact with previous works, we introduce an effective coupling parameter $V_0$ such that
(see~\cite{ThCl,V0def})
\bea
\varGamma(D^{*+}\to D^0\pi^+)&=&2V_0\frac{q_{0c}^3}{m_{\pi}^3},
\label{width1}\\
\varGamma(D^{*0}\to D^0\pi^0)&=&V_0\frac{q_{00}^3}{m_{\pi}^3}
\label{width00}
\eea
and, therefore,
\be
\frac{g_f^2}{4m_*^2}=\frac{6\pi V_0}{m_\pi^3},
\ee
where the mass difference between the charged and neutral states is neglected in the denominators.

The effective coupling $V_0$ can be estimated from the data on the
total $D^{*+}$ width \cite{PDG},
\be
\varGamma_c=(96\pm 22)~\mbox{keV},
\label{Dcwidth}
\ee
and the branching fractions \cite{PDG}
\bea
B(D^{*+}\to D^+\pi^0)&=&(30.7\pm 0.5)\%,
\label{brpi0}\\
B(D^{*+}\to D^0\pi^+)&=&(67.7\pm 0.5)\%.
\label{brpiplus}
\eea
The value $V_0=1.3$~MeV \cite{ThCl} complies well with the data on the $D^{*+}$ pionic decays.

The total $D^{*0}$ width is not known experimentally, however it can be estimated from
(\ref{width0}), (\ref{widthc}), (\ref{brpi0}), and with the help of the branching fraction $B(D^{*0} \to D^0
\pi^0)=(61.9\pm 2.9)\%$ \cite{PDG} to be
\be
\varGamma_0=(65\pm 15)~\mbox{keV}.
\label{D0width}
\ee

Summarising the discussion above, we can build the generic $D^*D\pi$ vertex in the form
\be
\veg(\veq)=g\veq\frac{\Lambda^2}{\Lambda^2+\veq^2},\quad g=\frac{\sqrt{6\pi V_0\vphantom{1^2}}}{m^{3/2}_{\pi}},
\label{ggg}
\ee
where $\veq$ is the relative momentum in the $D\pi$ system, and a form factor with the cut-off parameter $\Lambda$ is
introduced, as was explained above.

\begin{figure}[t!]
\centerline{\epsfig{file=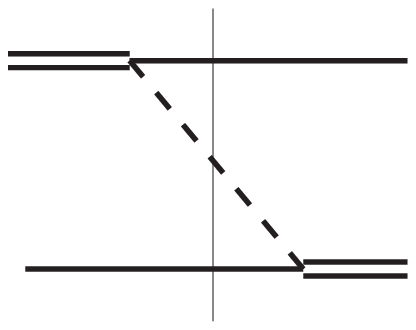, width=0.2\textwidth}\hspace*{5mm}\epsfig{file=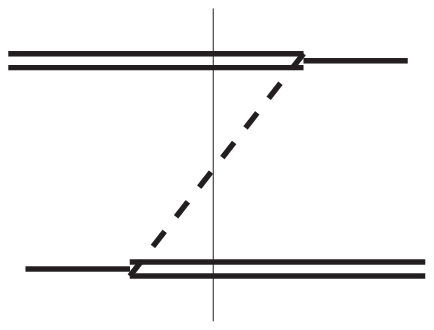, width=0.2\textwidth}}
\caption{{\bf Figure~\thefigure.}~Diagrams in the Time-Ordered Perturbation Theory responsible for the
potentials $V_1$ (left plot) and $V_2$ (right plot) from (\ref{VV1V2}). The double and single solid line are for
the $D^*$ and $D$, respectively, while the dashed line is for the pion. The thin vertical line pin-points the
intermediate
state.}\label{V1V2fig}
\end{figure}

The OPE potential $V$ in the coupled $(D^*\bar{D})$-$(\bar{D}^*D)$ system can now be built either
in the covariant formalism for the propagator of the intermediate pion or in the framework of the Time-Ordered
Perturbation Theory. In the latter case it is given by the sum of two possible orderings,
\be
V=V_1+V_2,
\label{VV1V2}
\ee
which correspond to the $D\bar{D}\pi$ and $D^*\bar{D}^*\pi$ intermediate state,
respectively (see Fig.~\ref{V1V2fig}). In what follows we consider the
$C$-even isosinglet channel.

The static limit of the potential (\ref{VV1V2}) can be obtained in the standard manner. If one neglects isospin
breaking due to the mass difference between charged and neutral $D^{(*)}$ mesons, then the covariant form of the static
potential in the momentum space is
\be
V_{\rm stat}^{nn'}(\veq)=-\frac{3}{(2\pi)^3}
\frac{g_n(\veq)g_{n'}(\veq)}{\veq^2+[m_{\pi}^2-(m-m_*)^2]},
\label{vstatic}
\ee
where the factor 3 accounts for the sum of the neutral- and charged-pion exchanges, the spin indices $n$ and $n'$
are to be contracted with polarisation vectors of the $D^*$ mesons, $m$ and $m_*$
are the masses of the $D$ and $D^*$ mesons, respectively (charged and neutral masses being indistinguishable in the
exact isospin-conserving limit). Notice that the potential (\ref{vstatic}) is two times smaller than the one used in
\cite{ThCl}, so that, effectively, the coupling parameter $V_0$ used in \cite{ThCl} is two times larger than
$V_0=1.3$~MeV compatible with the data
on the $D^*$ meson pionic decays. Thus we are forced to use the value $V_0=2.6$~MeV to reproduce
the results of \cite{ThCl}.

In the framework of the Time-Ordered Perturbation Theory expression (\ref{vstatic}) can be presented as
$$
V_{\rm stat}^{nn'}(\veq)=-\frac{3}{(2\pi)^3}g_n(\veq)g_{n'}(\veq)\Bigl(V_1^{\rm stat}(\veq)+V_2^{\rm stat}(\veq)\Bigr),
$$
\bea
V_1^{\rm stat}(\veq)&=&\frac{1}{2E_\pi(E_\pi+m-m_*)},
\label{first}\\
V_2^{\rm stat}(\veq)&=&\frac{1}{2E_\pi(E_\pi+m_*-m)},
\label{second}
\eea
where $E_\pi=\sqrt{\veq^2+m_{\pi}^2}$.

Since $m_* \approx m+m_{\pi}$, then one has $V_2 \ll V_1$ for the momenta $|\veq|$ of order a few hundred MeV. On the
other hand, as seen from expression (\ref{first}), the static approximation for the potential $V_1$ is inadequate, and
the full treatment of the $D\bar{D}\pi$ intermediate state is required.
\section{Formulation of the problem}

In~\cite{Baru:2011rs} the role of the dynamical pions played in a near-threshold resonance (at the example of
the $X(3872)$ charmonium) was discussed in detail. The key idea of the formalism used in~\cite{Baru:2011rs} was
to separate the short- and long-range interaction and to study in detail the role played by the long-range part of
the OPE. Meanwhile the short-range part of the OPE (together with other possible short-range interactions, such as, for
example, $s$-channel coupling to charmonium) was absorbed into a constant interaction described by a counter-term. For a
given
value of the cut-off $\Lambda$, the counter-term was tuned to guarantee the existence in the system of a
bound state with the given binding energy. As a result, the leading dependence of physical observables on $\Lambda$ was
absorbed by the counter-term. This approach, standard for effective theories, allowed one to use values of
the cut-off as small as just a few hundred MeV and therefore to deal with nonrelativistic kinematics for all particles
involved and to neglect the second ordering term as well as all contributions coming from higher multiparticle
intermediate states. On the contrary, the
aim of the present research is to investigate the OPE (its short-range part) from the point of view of the
possibility to bind the $D\bar{D}^*$ system alone. To this end, as was explained above, we consider the values of
$\Lambda\lesssim$ 1~GeV, as prompted by quark models. This implies that, while $D$ mesons can still be treated
nonrelativistically, we are forced to resort to relativistic kinematics for the pions. Technically this is the only
difference between the formulae used in this Letter and those found in~\cite{Baru:2011rs}, while their derivation
remains the same. We therefore skip here all details of the formalism (they can be found in~\cite{Baru:2011rs})
and only pin-point the essential differences between the two approaches.
\begin{figure}[t!]
\centerline{\epsfig{file=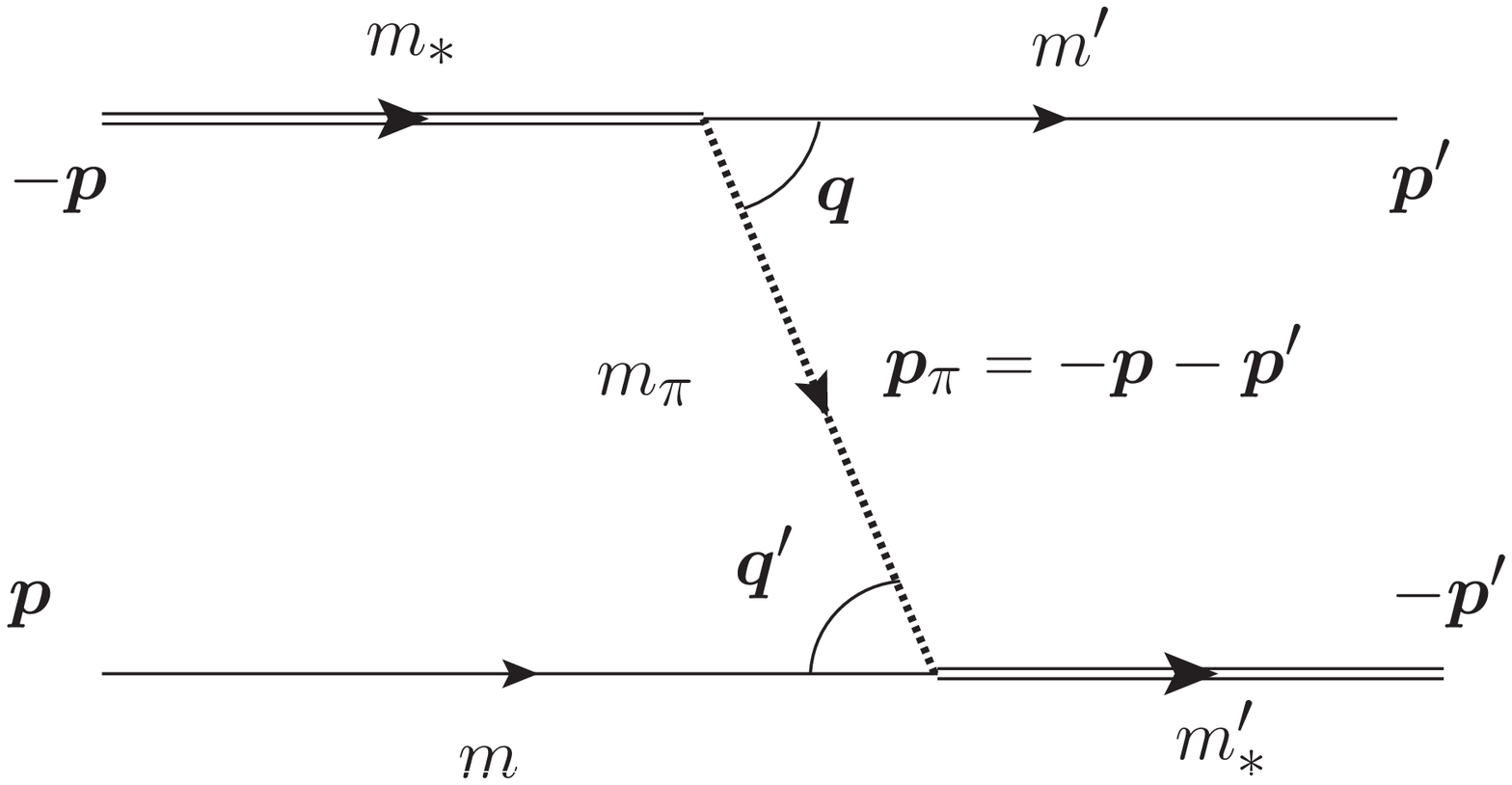, width=0.4\textwidth}}
\caption{{\bf Figure~\thefigure.}~Kinematics of the $D \bar D^*$ scattering due to the OPE. Double lines denote $D^*$'s
while single lines denote $D$'s.}\label{diagram}
\end{figure}
In particular, the OPE interaction graph depicted in Fig.~\ref{diagram} is described by the potential
\be
V^{nn'}_{ik}(\vep,\vep')=-\frac{1}{(2\pi)^3}\frac{g_n(\vep'+\alpha_{ik}\vep)g_{n'}
(\vep+\alpha_{ik}\vep')}{D_{3ik}
(\vep,\vep')},
\label{Vmn}
\ee
where the vertex function $\veg(\veq)$ is given in (\ref{ggg}) above, while $n$, $n'$ are $i$, $k$ are spin and
channel indices, respectively. The three-body propagator reads
\be
D_3(\vep,\vep')=2E_\pi(E_\pi-\mu-i0),
\label{D3}
\ee
where
\be
\mu=m_{*0}+m_0+E-\sqrt{m^2+p^2}-\sqrt{m'^2+p'^2}
\label{mu}
\ee
and
\be
E_\pi=\sqrt{(\vep+\vep')^2+m_\pi^2}.
\label{Epi}
\ee
The energy $E$ is defined with respect to the $D^0\bar{D}^{*0}$ threshold.
The coefficients $\alpha$ and $\alpha'$ can be extracted from the standard relativistic textbook formula for the
relative momentum written in terms of the single-particle momenta in the form (in notations of Fig.~\ref{diagram}; see,
for example,~\cite{novog,Kalashnikova:1996pu}):
\begin{eqnarray}
&\ds\alpha=\frac{1}{\sqrt{\varepsilon'^2-p^2}}\left[\sqrt{m'^2+p'^2}+\frac{\vep\vep'}
{\varepsilon'+\sqrt{\varepsilon'^2-p^2}}\right],&\\
&\ds\alpha'=\frac{1}{\sqrt{\varepsilon^2-p'^2}}\left[\sqrt{m^2+p^2}+\frac{\vep\vep'}{\varepsilon+\sqrt{
\varepsilon^2-p'^2}}\right],&
\end{eqnarray}
where
\be
\varepsilon=\sqrt{m^2+p^2}+E_\pi,\quad \varepsilon'=\sqrt{m'^2+p'^2}+E_\pi.
\ee
\begin{figure}[t]
\centerline{\epsfig{file=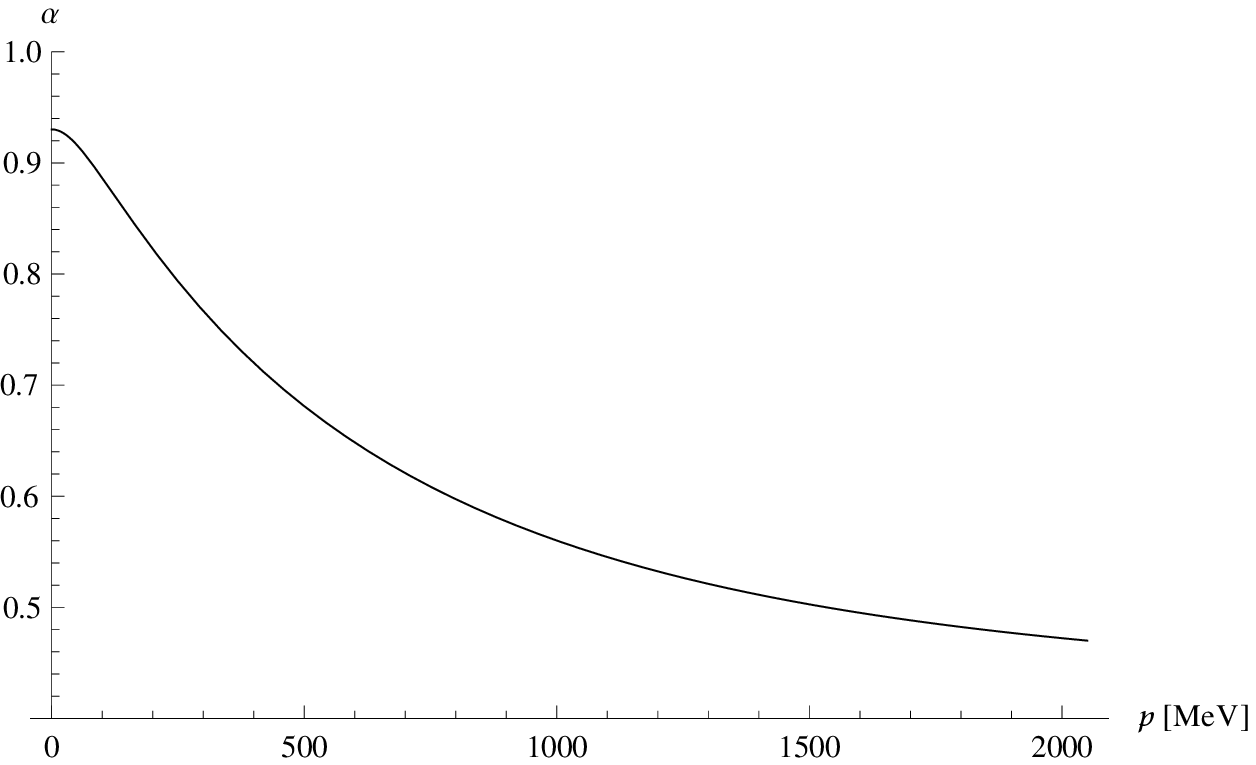, width=0.45\textwidth}}
\caption{{\bf Figure~\thefigure.}~Dependence of the coefficient $\alpha$ on the momentum $p$
($\vep'=\vep$).}\label{alphaplot}
\end{figure}
Nonvanishing components of the OPE potential depicted in Fig.~\ref{diagram} read:
\begin{eqnarray*}
V^{nn'}_{0 \bar 0}&=&V^{nn'}_{\bar 0 0},~m=m'=m_0,~m_{\pi}=m_{\pi^0},\label{V1}\\
V^{nn'}_{c \bar c}&=&V^{nn'}_{\bar c c},~m=m'=m_c,~m_{\pi}=m_{\pi^0},\\
V^{nn'}_{c \bar 0}&=&V^{nn'}_{\bar c 0},~m=m_0,~m'=m_c,~m_{\pi}=m_{\pi^c},\\
V^{nn'}_{0 \bar c}&=&V^{nn'}_{\bar 0 c},~m=m_c,~m'=m_0,~m_{\pi}=m_{\pi^c},\label{V4}
\end{eqnarray*}
where the channels are defined as:
\begin{eqnarray*}
&&|0\rangle=D^0\bar{D}^{*0},\quad|\bar{0}\rangle=\bar{D}^0 D^{*0},\\
&&|c\rangle=D^+ D^{*-},\quad|\bar{c}\rangle=D^- D^{*+}.
\end{eqnarray*}

The potentials $V^{nn'}_{ik}$ are to be supplied by the isospin factors:
\be
\langle 0|\vec\tau_1\cdot\vec\tau_2|\bar{0}\rangle=\langle c|\vec\tau_1\cdot\vec\tau_2|\bar{c}\rangle=1,
\ee
for the neutral pion exchange, and
\be
\langle 0|\vec\tau_1\cdot\vec\tau_2|\bar{c}\rangle=\langle c|\vec\tau_1\cdot\vec\tau_2|\bar{0}\rangle=2,
\ee
for the charged pion exchange, which we take into account explicitly in the Lippmann--Schwinger equations (see the
system (\ref{aa}) below).

The static approximation is obtained from expression (\ref{Vmn}) by setting $\alpha=\alpha'=1$ and
$\mu=m_{*0}+m_0-m-m'$. In Fig.~\ref{alphaplot} we plot
the dependence of the coefficient $\alpha$ on the momentum $p$. It is clearly seen from this plot that $\alpha$
decreases
fast with the increase of the momentum $p$ that, by virtue of (\ref{ggg}) and (\ref{Vmn}), leads to an effective
suppression of the OPE potential as compared to the static approximation with $\alpha=\alpha'=1$ for all momenta.
In this approximation, the potential is local, that is, it depends on $\veq=\vep+\vep'$ and, in the isospin-conserving
limit, the first ordering static potential (\ref{first}) is immediately reproduced.

Then we can write the system of coupled
Lippmann--Schwinger equations for the $D\bar{D}^*$ $t$-matrix elements $a_{00}^{nn'}(\vep,\vep')$ and
$a_{c0}^{nn'}(\vep,\vep')$ (see \cite{Baru:2011rs} for the details). In the $C$-even channel it reads
\bea
\left\{
\begin{array}{rl}
a_{00}^{nn'}(\vep,\vep',E)&\hspace*{-2mm}=V_{00}^{nn'}(\vep,\vep')\\[2mm]
&\hspace*{-10mm}-\ds\int\frac{d^3s}{\Delta_0(s)}V_{00}^{nm}(\vep,\ves)a_{00}^{mn'}(\ves,\vep',E)\\[2mm]
&\hspace*{-10mm}-\ds 2\int\frac{d^3s}{\Delta_c(s)}V_{0c}^{nm}(\vep,\ves)a_{c0}^{mn'}(\ves,\vep',E)\\[2mm]
a_{c0}^{nn'}(\vep,\vep',E)&\hspace*{-2mm}=2V_{c0}^{nn'}(\vep,\vep')\\[2mm]
&\hspace*{-10mm}-\ds 2\int\frac{d^3s}{\Delta_0(s)}V_{c0}^{nm}(\vep,\ves)a_{00}^{mn'}(\ves,\vep',E)\\[2mm]
&\hspace*{-10mm}-\ds \int\frac{d^3s}{\Delta_c(s)}V_{cc}^{nm}(\vep,\ves)a_{c0}^{mn'}(\ves,\vep',E).
\end{array}
\right.
\label{aa}
\eea

The inverse propagators $\Delta_0$ and $\Delta_c$ take the form
\bea
&&\ds\Delta_0(p)=\frac{p^2}{2\mu_0}-E-\frac{i}{2}\varGamma_0,\nonumber\\[-3mm]
\label{Deltas}\\[-3mm]
&&\ds\Delta_c(p)=\frac{p^2}{2\mu_c}+\delta-E-\frac{i}{2}\varGamma_c,\nonumber
\eea
where $\delta=m_{*c}+m_c-m_{*0}-m_0=8.08$~MeV, the reduced masses are defined as
$$
\mu_0=\frac{m_{*0}m_0}{m_{*0}+m_0},\quad
\mu_c=\frac{m_{*c}m_c}{m_{*c}+m_c},
$$
and the widths $\varGamma_0$ and $\varGamma_c$ are given in (\ref{D0width}) and (\ref{Dcwidth}), respectively.

Notice that both two-body propagators $\Delta_0(p)$ and $\Delta_c(p)$ from (\ref{Deltas}) as well as the three-body
propagator
$D_3(\vep,\vep')$ from (\ref{D3}) generate contributions to the imaginary part of the interaction which was omitted
in~\cite{ch,ThCl} and which we keep here to preserve unitarity.

For the channel with the quantum numbers $1^{++}$, we are interested only in the $S$ wave in the final state. We
therefore use the projectors
\be
T_{SS}^{nn'}=\frac{1}{4\pi}\delta_{nn'},\quad T_{DS}^{nn'}=\frac{1}{4\pi\sqrt{2}}
(\delta_{nn'}-3n_nn_{n'})
\ee
to find the $a_{ik}^{SS}$ and $a_{ik}^{DS}$ matrix elements of the amplitude, where the superscripts $S$ and $D$ denote
the $S$- and $D$-wave components:
$$
a_{ik}^{nn'}(\vep,\vep',E)=a_{ik}^{SS}(p,p',E)T_{SS}^{nn'}+a_{ik}^{DS}(p,p',E)T_{DS}^{nn'}.
$$

Finally, we calculate the differential production rate in the $D^0\bar{D}^0\pi^0$ channel. Further details as well
as an explicit formula for the production rate can be found in \cite{Baru:2011rs}. A near-threshold singularity (a bound
state in neglect of imaginary part of the potential), if exists, reveals itself as a below-threshold peak in the
production rate.

\begin{table*}[t]
\begin{center}
\begin{tabular}{|c|c|c|c|c|}
\hline
&$V_1^{\rm stat}$&$V_1^{\rm stat}+V_2^{\rm stat}$&$V_1$&$V_1+V_2^{\rm stat}$\\
\hline
$V_0$=1.3~MeV&2750&1650&3800&2100\\
\hline
$V_0$=2.6~MeV&1350&950&2400&1050\\
\hline
\end{tabular}
\caption{{\bf Table~\thetable.}~The minimal cut-off parameter $\Lambda_{\rm min}$ (in MeV) consistent with a bound state
in the
$D\bar{D}^*$ system, that is the cut-off that, for a given coupling parameter $V_0$, provides a bound state residing
exactly at the $D^0\bar{D}^{*0}$ threshold.}\label{table1}
\end{center}
\end{table*}

\section{Results and discussion}

The aim of the present research is to check whether the $D\bar{D}^*$ system interacting through the OPE possesses a
bound state near threshold. As the first
step, we fix phenomenologically adequate values $V_0=1.3$~MeV and $\Lambda=1$~GeV and find that
no bound state exists in the system, that we interpret as a demonstration that OPE is not sufficiently binding to be
responsible for the formation of the $X$. To have a better insight, we keep $V_0$ fixed and increase the cut-off until a
bound state appears at the threshold, that is with zero binding energy, that gives us the minimal value $\Lambda_{\rm
min}=3800$~MeV compatible with the existence of a bound state in the $D\bar{D}^*$ system. We repeat the
same calculation in the static limit, as explained before, and in neglect of the OPE imaginary part, to arrive
at the minimal cut-off $\Lambda_{\rm min}^{\rm stat}=2750$~MeV. The difference in the $\Lambda_{\rm min}$ between the
full treatment and the static approach must be ascribed to the effect of dynamical pions, inclusion of the imaginary
part of the OPE potential as well as to the momentum dependence of the coefficients $\alpha$. Notice that in the
calculation performed only the first ordering potential $V_1$ was kept. However, large values of the cut-off parameter
$\Lambda$ found indicate that large momenta (with $|\vep|\gg m_\pi$) may float in the $D$-meson loops, and therefore the
second ordering potential $V_2$ may give a sizable contribution. We therefore supply the OPE potential (\ref{Vmn}) with
an extra contribution, coming from the second ordering, for which we stick to the static form. Technically this amounts
to adding to $V_{ik}^{nn'}(\vep,\vep')$ from (\ref{Vmn}) a second term of the same form, however with $D^*$ and $D$
masses
interchanged and with the static limit applied, as explained above. As a result we find $\Lambda_{\rm min}=2100$~MeV and
$\Lambda_{\rm min}^{\rm stat}=1650$~MeV, respectively. For convenience we collect our results in Table~\ref{table1}.

One can draw two
conclusions from the results obtained. On one hand, even inclusion of the second ordering term and resorting to the
static limit does not bring us to the phenomenologically adequate values of the cut-off parameter $\Lambda$, the minimal
values required for the latter being essentially larger than 1~GeV. Besides that we observe that the contribution of the
second ordering appears to be quite significant both in the static limit and in the full treatment. In this situation
one should, in principle, go beyond the static approximation for the second ordering contribution too that, {\em inter
alia}, opens a Pandora box of intermediate states via the transition chain
\be
\bar {D}^*D^* \pi \leftrightarrow \bar {D} D^* \pi \pi \leftrightarrow \bar{D} D \pi\pi\pi\ldots,
\ee
undermining in such a way the naive few-body treatment of the problem.

Finally, we repeat the same calculations for $V_0=2.6$~MeV which, as
explained above, is the value of the coupling which was effectively used in \cite{ThCl}. For selfconsistency, in the
full calculation, the values of $\varGamma_0$ and $\varGamma_c$ used in the two-body propagators (\ref{Deltas}) were
also increased by a factor of 2. We reproduce the result of \cite{ThCl} if, in addition, the static
approximation is used, imaginary parts of the OPE are neglected, and both orderings are taken into account. Namely, a
bound state at the threshold appears for relatively small values of the cut-off, $\Lambda_{\rm min}^{\rm stat}\sim$
1~GeV (see Table~\ref{table1}). However, if the full form for the first ordering potential is used,
the value of $\Lambda_{\rm min}=2400$~MeV is required for the case of the first ordering potential alone, while
inclusion
of the second ordering in the static form lowers the value of the cut-off to phenomnologically accepted one,
$\Lambda_{\rm min}=1050$~MeV. This is to be confronted with $\Lambda_{\rm min}=2100$~MeV obtained for the full
treatment using the value $V_0=1.3$~MeV compatible with the data on $D^*$ decays. We are led to conclude therefore that
one-pion exchange is not sufficiently binding to produce the $X(3872)$ state and that other short-range dynamics are
responsible for its formation.
\smallskip

This work is supported in part by the DFG and the NSFC through funds provided to the sino-germen CRC 110 ``Symmetries
and the Emergence of Structure in QCD''. A.N. would like to acknowledge hospitality of the staff of IKP,
Forschungszentrum J{\"u}lich, where part of this work was done.

\end{document}